# Early Telescopes and Ancient Scientific Instruments in the Paintings of Jan Brueghel the Elder

Pierluigi Selvelli and Paolo Molaro
INAF, Osservatorio Astronomico di Trieste, Italy

**Abstract:**

Ancient instruments of high interest for research on the origin and diffusion of early scientific devices in the late XVI – early XVII centuries are reproduced in three paintings by Jan Brueghel the Elder. We investigated the nature and the origin of these instruments, in particular the spyglass depicted in a painting dated 1609-1612 that represents the most ancient reproduction of an early spyglass, and the two sophisticated spyglasses with draw tubes that are reproduced in two paintings, dated 1617-1618. We suggest that these two instruments may represent early examples of keplerian telescopes. Concerning the other scientific instruments, namely an astrolabe, an armillary sphere, a nocturnal, a proportional compass, surveying instruments, a Mordente's compass, a theodolite, etc., we point out that most of them may be associated with Michiel Coignet, cosmographer and instrument maker at the Court of the Archduke Albert VII of Hapsburg in Brussels.

# 1. The Historical and Cultural Context

The reign of the Archdukes Albert VII of Hapsburg-Austria and Isabella Clara Eugenia of Hapsburg-Spain, (1596-1625), certainly represented a key period of the history of the Low Countries. The Archdukes ruled the (catholic) South, while the (calvinist) North Provinces were governed by the regents, the Princes Maurice and Henry of Nassau-Orange, who represented the wealthy merchant class. The government of the Archdukes was characterized by the first period of peace, the Twelve Years Truce, signed in Antwerp on 9 April 1609 after almost four decades of war. This stimulated economical growth and the parallel renaissance of culture and science under the active support of Albert VII, as manifested by his patronage of artist like P.P. Rubens and J. Brueghel, architects like Coudemberg, scientist like the mathematician, engineer, and geographer Michiel Coignet, and humanists like the philologist Justus Lipsius. Jan Brueghel was the second son of of Pieter Bruegel the Elder. He traveled to Italy around 1590 and worked in Rome, Naples and Milan, at the court of Cardinal Borromeo, his first patron, before reaching Antwerp in 1596 were he was a *protege* of the Archdukes, In 1610 he became their *court painter* .

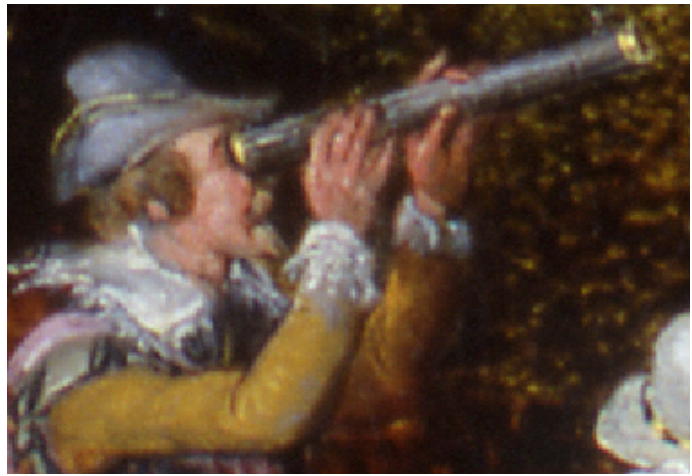

Fig. 1 A detail of the painting *Extensive Landscape with View of the Castle of Mariemont* by J. Brueghel the Elder, ca. 1609-1612. Virginia Museum of Fine Arts, Richmond. The Adolph D. and Wilkins C. Williams Fund. Photo: Ron Jennings. Note the Archduke Albert VII who looks through a spyglass. This is the most ancient reproduction of this instrument and represents one of the first spyglasses ever built.

# 2. The Early Spyglass in the Painting *Extensive Landscape with View of the Castle of Mariemont*

The painting *Extensive Landscape with View of the Castle of Mariemont* is exhibited at the Virginia Museum of Fine Arts (VMFA) in Richmond, VA, USA. Mariemont Castle, located near

Brussels, had restoration for several years, and a study of the development of the building by comparison with paintings of ascertained dates, indicated that the date of the painting completion was in the years 1608 - 1611. In the central part of the painting one can see the important detail of the archduke Albert VII who looks through a spyglass. The instrument has a cylindrical shape and metallic appearance, probably made of tin, with gilded rings on both end sides (Fig. 1). By comparison with other figures and objects one can estimate that its length is around 40-45 cm and its diameter around 5 cm.

This painting represents the most ancient reproduction of a spyglass and we guess that the depicted instrument actually represents one of the first spyglasses ever built. Until now the painting by Jusepe de Ribera (1614) has been recognized as the earliest. The presence of the spyglass in the painting can be used to exclude the year 1608 in the range of the possible dates for its completion, since the first spyglasses appeared in Autumn 1608, while the trees of the landscape and other details in the painting indicate that the season is (late) summer.

## 3. The Spyglasses with Draw Tubes in the Paintings of El Prado .

The painting of Jan Brueghel and P. Rubens *La Vista* and the painting *El Olfato y la Vista* by J. Brueghel and other painters, both conserved at the Museum El Prado in Madrid, belong to a bedini of paintings known as *Los cinco sentidos* (also named *The Allegory of the Five Senses* ).
The first painting (oil on wood) is drawn with great accuracy and depicts a hall in the ancient royal Palace of Brussels on the hill of Coudemberg, residence of the Archdukes, where they kept their collection of paintings, precious items, and scientific instruments, most of which related to astronomy,. The painting was completed by 1617, as demonstrated by the date that appears on a roll of several papers lying over a book in the lower part of the painting, besides the author's signature.
The second painting (oil on canvas) was completed around 1618. It was commissioned by the City of Antwerp to J. Brueghel and other painters, to celebrate the visit of the Archdukes to the City. The painting, which actually is a copy of the original that was lost in the fire of the Castle of Coudemberg near Brussels in 1731, includes also several instruments that are very similar to those reproduced in the previous painting. For details about the quite intrigued origin and story of the two paintings we refer to the studies of Bedini (1971) and Diaz-Padron (1992).

The telescope that appears in the first painting (Fig. 2), between the figures of Venus and Cupid, consists of a main tube and seven draw-tubes, all of which appear to be made of metal (probably silver). Each of the intermediary draw tubes terminates in an enlarged collar that appears to be made of the same metal. The lenses are housed in large rounded terminals. The instrument is fixed into a curved metal sleeve support attached to a brass joint which can be adjusted for angle. The pedestal consists of a turned column terminating in a simple saucer-shaped round base. A comparison with other objects depicted in the painting indicates a maximum and minimum width of the draw tubes of about 7.5 and 2.5 cm, respectively, and an estimated total length, if the tubes were all drawn, close to 170 cm.

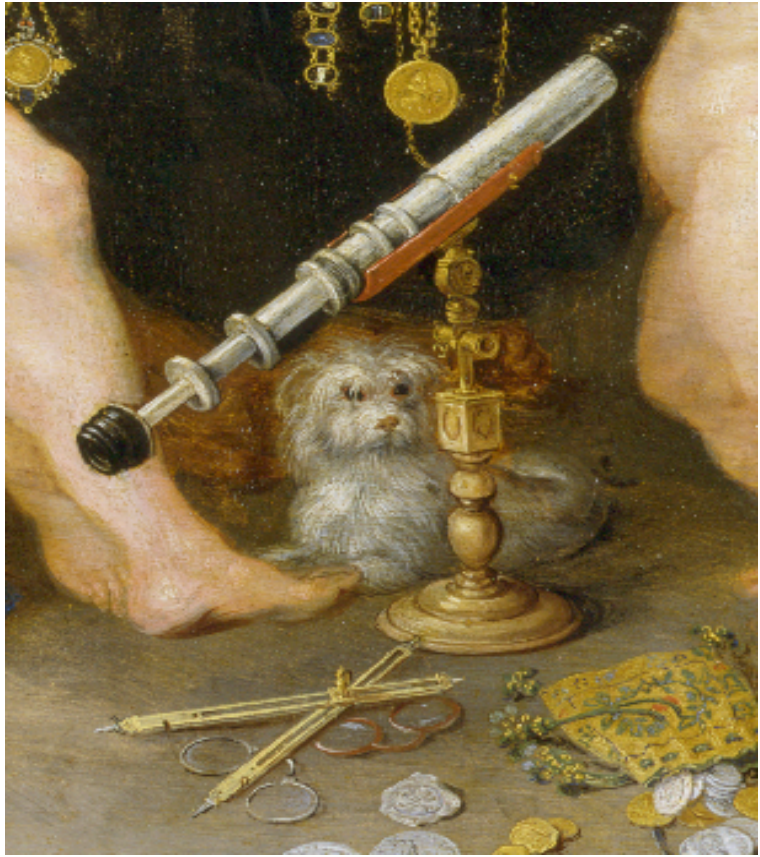

Fig.2 A detail of the painting *La Vista* by J. Bruegel and P.P. Rubens, 1617. Madrid, Museo Nacional del Prado. Note the spyglass with several draw tubes, quite sophisticated for the epoch and, on the floor, a proportional compass by M. Coignet

A similar telescope is reproduced also in the second, larger painting (Fig. 3). The main difference between the two telescopes is in the number of draw tubes (eight instead of seven) and in the color of the rings (black instead of silvery). Also the pedestals are different (in the two cases). Thus, despite the overall similarity of their form, which clearly indicates the same origin (craftsman), the two telescopes are, apparently, two separate instruments. We note that Bedini (1971) described this telescope as constructed with cardboard tubes covered with white or light-colored vellum. This seems hardly compatible with the metallic aspect of the tubes, as suggested by their color and reflectivity.

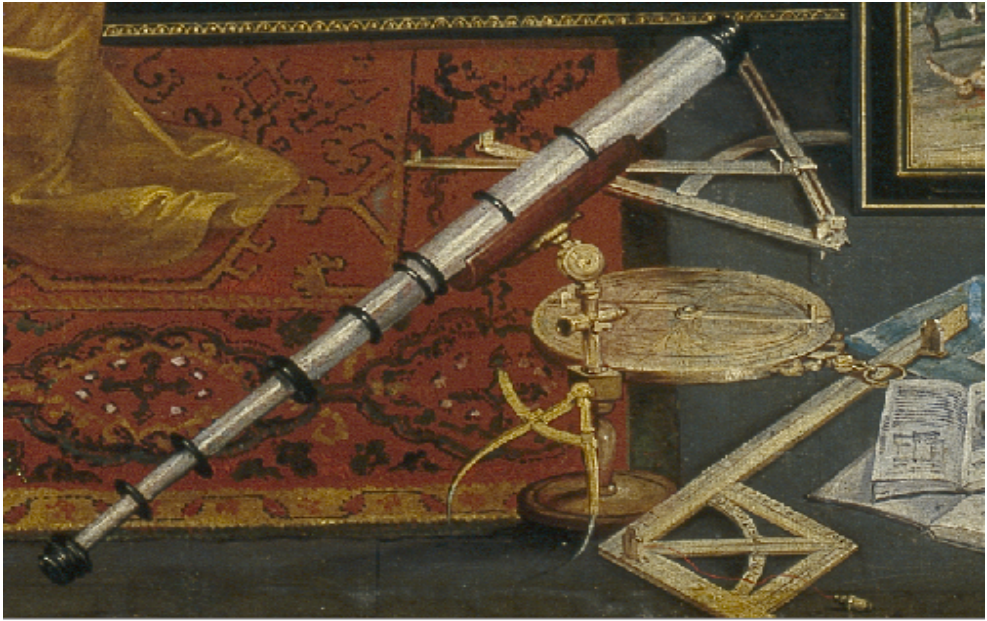

Fig.3 A detail of the painting *La vista y el olfato* by J. Bruegel et al., ca. 1618. Madrid, Museo Nacional del Prado. Note a second spyglass with draw tubes, similar but not identical to that represented in the first painting (Fig. 2). Note also an astrolabe, a caliper, and a gunner's rule; these three instruments appear identical to those depicted in the first painting.

## 4. Early keplerian Telescopes?

The two telescopes appear surprisingly sophisticated if compared to other telescopes of the same epoch, and their metallic nature is quite unusual in early spyglasses. A definite upper limit for the manufacturing of the instrument depicted in Fig. 2. is set by the completion of the paintings in 1617 and 1618, respectively. What is quite surprising is that there is no record of similar instruments until about three decades later (cf. Van Helden, 1999). After examining illustrations of early telescopes, we conclude that the closest resemblance in shape and length occurs with the illustrations reported in C. Scheiner's works (Disquisitiones Mathematicae, 1614, and Rosa Ursina , 1631). We suggest that these instruments may represent early examples of keplerian telescopes, which makes their presence even more striking. This supposition is based on three considerations (see Selvelli and Molaro 2009, and Molaro and Selvelli 2010 for additional details):

1. the presence of a quite large terminal (eyepiece) seems incompatible with a Galilean (Dutch) mounting. In this case, the negative lens needs the eye to be brought as close as possible so that the eye's pupil becomes the aperture stop and the exit pupil. A careful inspection of the terminal has not revealed any evidence for the presence of a lens.

2. the overall length of the telescope is estimated to be about 170 cm. Even with low powers this would imply such a small field of view to make the spyglass quite useless.

3. The width of the tubes nearer to the eyepiece is smaller than that in the (few) known examples of a *dutch* spyglass with draw tubes. This characteristic is hardly compatible with a *dutch* configuration, because, again, it would result in a very small field of view.

We have quantitatively verified these three considerations by using the Optical Ray Tracer, a software for the simulation of optical systems, for various geometrical and optical configurations which are compatible with those of the spyglasses in the two paintings, i.e. total length ~ 175 cm., objective lens diameter ~ 6.0 cm., ocular lens diameter ~ 2.5 cm, magnification ~ 10-20 times. We note that the origin and development of the *astronomical* telescope, consisting of two convex lenses, is uncertain and open to question. It was theoretically described by Kepler in his *Dioptrice* of 1611 but it is not clear when the first *astronomical* telescope was manufactured. *Rosa Ursina*, completed by Scheiner in 1630, is the first book containing a reference to an astronomical telescope.

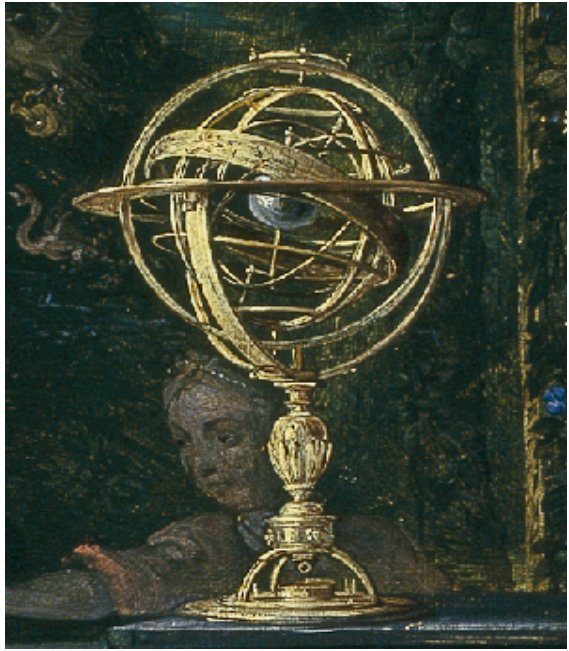

Fig.4 Same as in Fig. 2. An armillary sphere; craftsman is uncertain, see text.

## 5. The Archduke Albert VII and the Spyglasses

Several documents demonstrate the strong interest of the Archduke for the newly designed instrument:

1. According to Antonius Maria Schyrley de Rheita, the Marquis Ambrogio Spinola, Commander of the Spanish Army in the Flanders, bought a spyglass in The Hague towards the end of 1608, probably made by Lipperhey, and offered it to Archduke Albert (Van Helden, 1977). Spinola was in The Hague in that period, as delegate of the Spanish Governor for peace negotiations with Maurice of Nassau, Staatshouder of the Seven Provinces (Van Helden, 1977).

2. G. Bentivoglio, *nuncio* of the Pope in Brussels, in a letter of April 2 1609 to Cardinal Borghese wrote: *When the marquis Spinola returned from Holland, the archduke was most desirous to obtain such an instrument and indeed it came about that one came into their hands*

3. The noble man Daniello Antonini, from Udine, Italy, pupil and friend of Galileo, during his stay at the court of the Archduke, in a letter to Galileo dated September 1611 writes: *I have seen the best spyglasses that have been built here, including those of the first inventor*. We know from Daniello Antonini, who was serving with the Archduke in Brussels, that Albert VII owned copies of early spyglasses. Antonini, on September 2, 1611, wrote a letter to his former master Galileo (Bibl. Naz. Fir., Mss. Gal., P. VI, T. VIII, car.37) informing him that the Archduke had obtained some spyglasses from the *first inventor*, although of lower quality in comparison to those owned by Galileo. Thus, it is likely that the *tube* held by the Archduke represents one of these spyglasses mentioned by Antonini.

4. Pierre Borel, in his *De vero telescopii inventore* writes: *our artisan [Sacharias Janssen] first made tubes of 16 inches, and offered the best ones to Prince Maurice and the Archduke* (Van Helden 1977).

5. A document conserved in the Royal Archives of Belgium (Houzeau, 1883) shows that on 5 May 1609 the Archduke in Bruxelles ordered the payment of 90 Brabant florins to the royal goldsmith for *two tubes to see at a distance*.

6. In 1655 Willem Boreel, ambassador of the United Provinces of Netherlands in France, in a testimony to Pierre Borel declared *I have often heard that Sacharias Jansen and his father (Hans) first invented the microscope. This instrument was then presented to the Archduke Albert of Austria, who, in turn, presented it to the inventor Cornelis Drebbel, a good friend of mine* (Van Helden, 1977). This also indicates the existence of a link between the Archduke and Sacharias Jansen.

7. Concerning the alleged keplerian configuration of the spyglasses with draw-tubes, we note that the Jesuit Christopher Scheiner, active at Ingolstadt, Innsbruck and in Rome, in his *Rosa Ursina* of 1631 claimed that he made a keplerian instrument in 1614-1617 and showed it to the Archduke Maximilian III of Tirol, brother of Albert VII. In a letter of January 1615 he mentions a *new* instrument. According to documents in the Tyrolean State Museum studied by F. Daxecker (2004) Maximilian received a keplerian instrument in May 1616, and Scheiner added a third lens, thus manufacturing a terrestrial keplerian telescope. It seems plausible that Maximilian informed his brother Albert of this new instrument.

8. Also, Willem Boreel, ambassador of the United Netherlands in France, in a testimony to Pierre Borel in 1655, declared that in Middelburg Sacharias Jansen invented *the long sidereal telescopes, with which we inspect the other planets, the stars and the constellations* (Van Helden 1977). We guess that these *long sidereal telescopes* were of keplerian design.

# 6. The Archduke Albert VII and the Instrument Maker Michiel Coignet

Besides the spyglasses, the two paintings conserved at the El Prado Museum exhibit a notable collection of measuring devices, related to scientific and technical disciplines, e.g. astronomy, geometry, geography, cartography, navigation, architecture, and military engineering. Some of the instruments depicted in the two paintings were already briefly described by Belloni (1964) and Bedini (1971).The interest of the Archduke for these instruments is indicated by the dedications he received for books related to these disciplines. From them, we mention the book *Libro de instrumentos nuevos de geometria* by Don Andres Garcia de Cespedes, *Cosmografo mayor* of the King of Spain, in which the geometrical grounds of an instrument for leveling (the *Nivel*) are described, and the book *Globus Terrestris Planisphericus* written in Antwerp in 1613 by Ottavio Pisani, Galileo's correspondent.

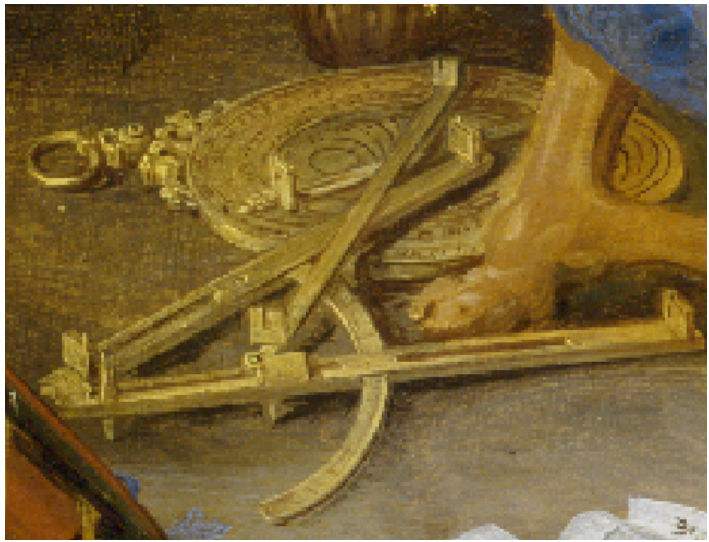

Fig.5  Same as in Fig. 2. An astrolabe by Coignet, and a folding rule.

We note that most of the instruments represented in the two paintings are in direct or indirect relationship with the instrument maker Michiel Coignet (1549-1623). This conclusion derives either from direct evidence, i.e., from the comparison with very similar instruments conserved in museums and signed by or attributed to Coignet, or from circumstantial evidence, based on the description of instruments made by Coignet in the literature.
Native of Antwerp, Michiel Coignet was the son of the goldsmith and instrument maker Gillis Coignet, and friend of Gerard Mercator and Gualterus Arsenius. (1512-1594). Coignet's family was leader in the fabrication of scientific instruments, including astronomical devices like armillary spheres, astrolabes, proportional compasses and navigation instruments. Michel Coignet started his career in 1568 as a school master and in 1572/1573 he was appointed as the city's wine gauger

(Meskens, 2002).In 1590 Coignet became Master of the Guild of Goldsmiths, and in 1596 entered the service of the archduke Albert VII as court mathematician and military engineer.  Coignet exceptional standing was recognized by Albert VII who titled him as court *cosmographe*  in 1604 .

   A list of scientific instruments that were made or studied by Michiel Coignet  includes clocks for the city of Antwerp, astrolabes conserved at the Boerhaave Museum in Leiden and at the  *Museo Maritimo* in Madrid, the circumferenter or Dutch Circle,  the cross staff improved with three transverse pieces, the nautical hemisphere, the nocturnal, described already in 1581, with a specimen of 1598 in Oxford, the sundial, the sector or proportional compass, for which Coignet is recognized as one of the inventors, the pantometric rule, and so on. Many of these instruments are present in the two paintings and it is likely that Coignet was indeed their maker or the curator of the collection at the Court of the Archduke, where he was appointed as court cartographer and engineer.
Coignet also  described  in manuscripts and books, e.g. *Traite' des Sinus*  1610,  *De Regulae pantometrae*  1612 and  *El uso del compas proportional*  1618, dedicated to the Archduke Albert VII, the use and applications of mathematical instruments like the proportional rule (reigle platte), the flat-legged proportional compass and the four-points proportional compass. In his book  *La geometrie* published in Paris in 1626 after his death, Michiel Coignet described several applications of both the *pantometre ou compas de proportion*  and the compass with eight points (Mordente's compass) to the solution of geometrical problems .

   As reported by  Favaro (1908), Coignet was in correspondence with Galileo. On March 31, 1588 Coignet wrote a letter to the young Galileo asking him for some information about his studies on the centroids of paraboloids. It was Ortelius that during a stay in Rome heard of the Galileo's mathematical studies and then informed Coignet. In this context, it is not surprising that Coignet is mentioned as a valued mathematician in a letter that Daniello Antonini, former pupil of Galileo in Padua, wrote on July 9, 1611 to Galileo during his stay in Bruxelles at the court of the archdukes.

## 7. Other Scientific  Instruments in the Paintings

   By a detailed examination of the two paintings exhibited at  *El Prado*  (hereinafter  A   for  *La Vista*  and  B   for   *la Vista y el Olfato* ), besides the  spyglasses, we detected the presence of 15 different scientific instruments, without mentioning the spectacles, the magnifying lenses, and the terrestrial globe. Most instruments are common to the two paintings, but a theodolite, a reduction compass, and a divider are in A only (Fig . 2 and 6), while another divider and a folding rule (sector ?) are in B only (Fig. 7). It is unfortunate and regrettable that in the literature there is no definite and commonly accepted terminology for scientific instruments.. While it is still acceptable that different names may be assigned to a given instrument, it is more disturbing that often the same name (terminology) is adopted to designate different instruments. This is especially true in the case of the instrument that in English is called *sector* , but is also called *reduction compass* or *proportional compass* . To make things worse, different authors mention different instruments with the same name (e.g. proportional compass). In addition, we note that what is called *proportional compass* in English, is called *compas de redution* in French, and similar terminology is found in Dutch, German, and Italian. This fact gave origin to much historical confusion and misunderstanding. See, in particular, the considerations by Drake (1978) and Meskens (1997).

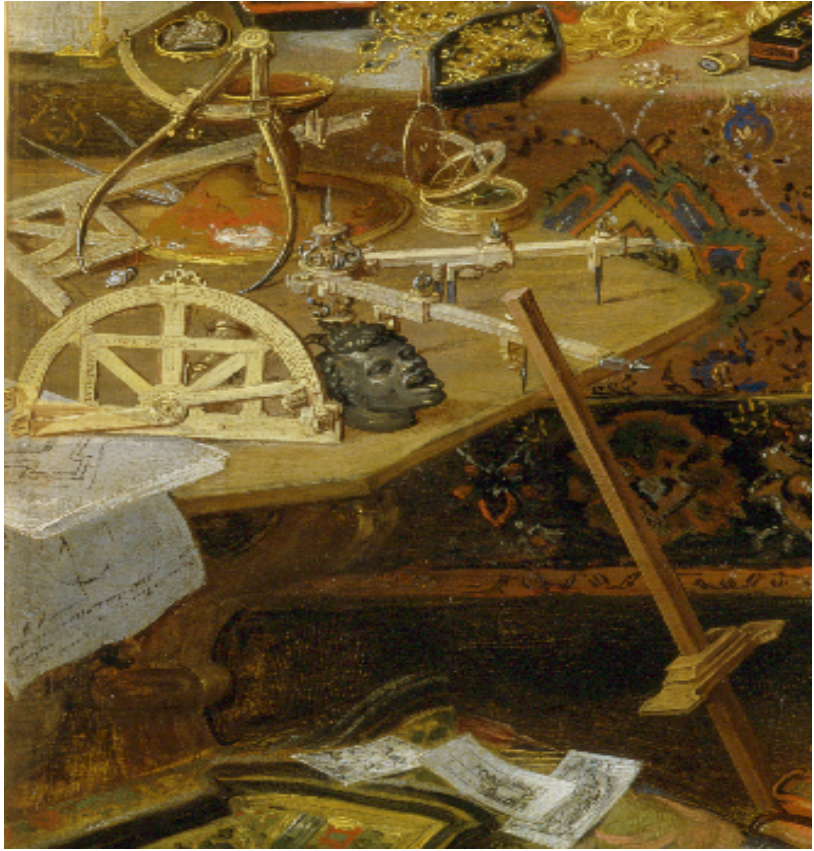

Fig.6 Same as in Fig. 2. Note the presence of seven scientific instruments; from left to right and from top to bottom: a divider , a gunner's rule, a caliper, a nocturnal (by. M. Coignet), a Mordente's compass, a theodolite, and a cross-staff.

Here follows the description of the various instruments displayed in the paintings. For a homogeneous comparisons, we use the terminology adopted for similar instruments in the Epact Electronic Catalogue of Scientific Instruments from European Museums (http://www.mhs.ox.ac.uk/epact), hereinafter EPACT , and in the Multimedia Catalogue of the

Institute and Museum of the History of Science, Florence (http://brunelleschi.imss.fi.it/museum), hereinafter IMMS ; alternative names are given in parenthesis.

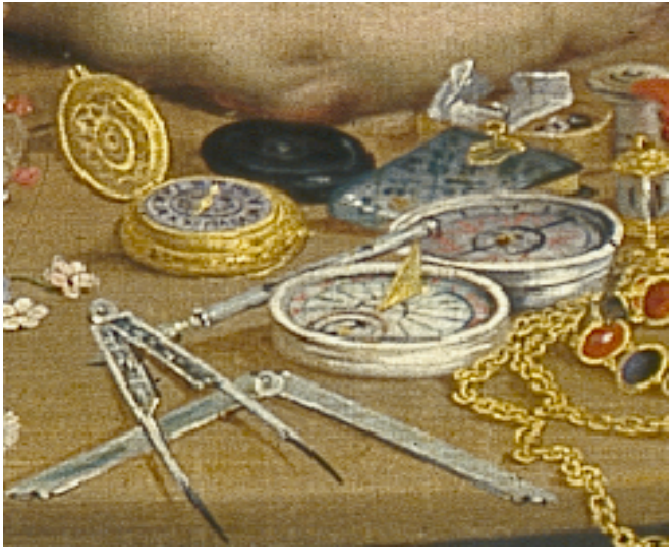

Fig.7 Same as in Fig. 3. An horizontal dial, a divider, and, below, a folding rule (sector ?).

***ARMILLARY SPHERE*** (Fig. 4) It is quite similar to the armillary sphere made by Giovanni Paolo Ferreri, (EPACT no.2), conserved at the Museum Boerhaave, Leiden. There is some resemblance also to that signed by Philippe Danfrie around 1570 (EPACT no. 6), but the dimensions of this sphere are much smaller (~ 20 cm only)

***ASTROLABE*** (Figs. 3 and 5) An astrolabe is present in each of the two paintings. Its design and several details clearly indicate that the origin is M. Coignet's workshop. This is evident from the comparison with the astrolabes conserved at the Museum Boerhaave (EPACT no. 73) and at the *Museo Naval* , of Madrid, both made by M. Coignet, the first dated 1601 (Antwerp), the second dated 1598. We note that Coignet learned the techniques of astrolabe construction from Gualterus Arsenius, instrument maker in Louvain, who was famous for his sophisticated astrolabes as well as armillary spheres, cross-staffs and astronomical rings.

***CALIPER*** (Fig. 3 and 6) This instrument was used to measure thickness and sizes, in particular the diameters of cannonballs. In the paintings it is close to the gunner's rule (see below).

***COMPASSES*** (dividers) (Figs .6 and 7) In painting A it is partially covered by the gunner's rule, in painting B it is to the left of the horizontal dial. Although in its various permutations this instrument was used as a compass, as a divider, or as a caliper, in this specific case the absence of insertable accessories seems to indicate that the two instruments are indeed dividers used for geometrical operations, e.g. to transfer measurements from one part of a drawing to another.

***CROSS STAFF*** (cross rod) (Fig. 6) The two orthogonal rods represent the base and the height of a triangle in which the visual rays of the observer form the other sides. Since the cross-rod can slide along the longer rod, the instrument allows the measurement of terrestrial (marine) and celestial angular separations, using the graduation on the longer rod. In 1581 Coignet suggested an improvement in the cross-staff obtained by the use of three transverse rods with three different graduated scales.

***DIVIDER*** (single handed divider) Near the base of the terrestrial globe. Similar to IMMS I.95. Dividers are one of the basic types of mathematical instruments In this case it is used is for taking off and comparing different distances.

***FOLDING RULE*** (proportional compass, sector ?) (Figs. 3 and 5 ) In this instrument, in addition to the two folding legs of the compass, one can note a semicircular sector and a cross arm, each bearing some measurement scales. After a detailed perusal in the on-line catalogs of both EPACT and IMSS, we found some partial resemblance with the proportional compasses made by Jost Miller dated 1616 (IMSS II.37) and with the instruments named folding rule (EPACT 274 and 276) of an unsigned maker, late XVI century. Instead, a closer resemblance comes out with Hood's sector depicted in the engraving by Whitwell in L'E.Turner (2001).

***GUNNER'S RULE*** (Galileo's squadra, elevation gauge, gunner's quadrant, folding rule) (Fig. 3 and 6) We note that one leg is longer than the other and the presence of a red string with a bob, that allowed to use the instrument as a plumb rule for leveling and regulating the elevation of a cannon, using a graduated quadrant arc. It is similar to Cole's Military sector, to Cole's Gunner's Folding Rule, (EPACT no. 286), to Tartaglia's elevation gauge (see also Drake, 1978) and to the Gunner's Rule of Hans Christoph Schissler (EPACT no. 290). A nearly identical instrument is reproduced in the *Portrait of Father Jean-Charles della Faille* by Anthony van Dyck, but is named *a quadrant and a plumb line* in the on-line description of the painting. In the same description another instrument in the painting, a sector (see below), is reported as *from the workshop of the Anwerp instrument maker Michiel Coignet* . It is likely that the Gunners rule may also be associated to M. Coignet. We note that, according to Meskens,
 (1997) in the manuscripts conserved at the Stadsbibliotheekt, of Antwerp (B264708), Coignet describes an instrument he called Tartaglia's squadra, probably a gunner's rule.

***HORIZONTAL DIAL*** (sundial) (Fig. 7). The depicted instrument bears a compass which allows the instrument to be oriented Northwards, and a gnomon. We found little resemblance with similar instruments in the on-line catalogs (just marginally with EPACT no. 328), none of which is attributed to Coignet. Therefore, the possibility of Coignet being the maker cannot be disregarded.

***MORDENTE'S REDUCTION COMPASS*** (Fig. 6) This instrument was first made around 1567 by Fabrizio Mordente (1532-1608), mathematician of Alexander Farnese, duke of Parma and Piacenza, as a drafting instrument, probably following some hints by Federico Commandino and its pupil, the Marquis Guidobaldo Del Monte. It consists of a pair of dividers (arms) with the addition of a number of sharp points that run on cursors that can slid along the arms and can be set to divide the arms in any desired proportion. This configuration provides a device that (with the aid of a rule) can establish the proportions between lines, geometrical figures, and solid bodies. Near the end of the XVI century, this instrument attracted the interest of scientists in the courts of Europe. We note that Mordente made friends with Coignet, who contributed to the spread of the instrument to various European courts . Two manuscripts with instructions by Coignet about the use of Mordente's compass appeared in 1608 (Meskens, 1997). In the full title of Coignet's work *La Geometrie* published in Paris in 1626, the use of the eight points compass by Mordente is explicitly mentioned. Coignet also suggested some improvements in the scale of the rule to be used with the compass by adding new lines for the representation of trigonometrical and geometrical relations.

The instrument represented in the painting is quite similar to that reproduced in Rose (1968) and conserved at the Adler Planetarium, and to that reproduced in Fig.6 of Meskens (1997), an illustration of Mordente's compass by Coignet, which is conserved at the Biblioteca Estense in Modena, Italy. Actually, the compass in the painting is very similar if not identical to that reproduced as Fig. 2 in Meskens (1997) and to that reproduced in Fig 7a of Bonelli's study (1959) about Mordente's work *La quadratura del cerchio* published in 1591 in Antwerp. We note that the description by Rose (1968) of an elaborate type of Mordente's compass which became widespread in the 1580's, *with cursors which fitted around each leg (instead of two cursors set into channels in the legs), the end points attached to cursors, thus making a total of four cursors* , fits very well with the depicted instrument.

*NOCTURNAL* (Astronomical compendium, aspectarium, equinoctial dial) (Fig. 6) In the two paintings, near Mordente's compass, there is a nocturnal. This identification, is based on similarities with catalogued instruments (EPACT 147, 265, 268, 365). Thanks to a recent study by L'E. Turner (2001) we can confidently identify M. Coignet as the maker, and the instrument as a nocturnal. The instrument in the paintings bears a strict similarity to the gilt-brass real one, signed Coignet 1602.

*PROPORTIONAL COMPASS* (?) (folding rule ?). (Fig.7) The identification of this instrument, that is present only in painting B, is made difficult by the small size of the object in the painting. We magnified this detail as much as possible in the special digital image obtained from *El Prado* Museum, without being able to detect the presence of any graduated scale in the legs. In any case, from a comparison with similar instruments in the on-line data-bank, we found quite close resemblance with IMSS I.92, I.93, I.105, and also with III.23. Comparison with other illustrations suggests also similarity with the sector of the Guidobaldo type. We did hope to find a positive identification of this instrument as a sector, an instrument to which both Coignet and Galileo gave practical and theoretical contributions; however this possibility, although attractive, remains unsubstantiated. It is well known (see Favaro 1908, Drake 1978, and Meskens 1997) that Coignet developed a pair of sectors which bear a close resemblance to Galileo's sector, and that in his manuscripts he mentioned he made the first sector for the Archduke Albert.

*REDUCTION COMPASS* (Fig. 2) According to Meskens (1997) who, however, names it *proportional compass*: *the instrument consists of two slotted arms with sharp points at each end. In the slots of the arms there is a cursor that serves as a pivot, and which can be held at the desired place by a screw. The pivot keeps the distances between the two points at each end in the same ratio, whatever their separation. It was a development from the X-shaped draftsman's instruments, which were fixed at ratios 2:1, 3:1, and so on* .

The instrument in the painting is very similar to that reproduced in Fig.1 of Mesken's (1997) study that also contains a description of the compass made by Coignet. We note that the instrument in the painting would instead indicate an earlier date, since the painting was completed in 1617.

From a detailed inspection of the on-line catalogs of ancient instruments we found that the only instrument that bears some resemblance with that in the painting is IMSS I-101, of an unknown maker .

*THEODOLITE* (Graphometer) (Fig. 6) This instrument was used for measuring angles in a half-circle scale. The angle is indicated by the position of the alidade on a stationary scale. The graphometer was developed by Philippe Danfrie around 1597. The instrument in the painting, in which one can note the inscriptions *latus rectum* and *latus versum* on the two sides of the inscribed rectangle, seems quite similar to the theodolite designed by Leonard Digges and to the graphometer in

IMSS I.78. We note that according to Meskens (1997), in the manuscripts conserved at the Stadsbibliotheekt, Antwerp (B264708), Coignet also described a kind of theodolite called diopter.

## 9. Conclusions

The spyglass depicted in a detail of the painting *Extensive Landscape* represents the most ancient reproduction of such an instrument and actually it may represent one of the first spyglasses ever built; until now the painting by Jusepe de Ribera (1614) has been recognized as the earliest. The two spyglasses with draw tubes, reproduced in two paintings conserved at *El Prado* and dated 1617-1718, are surprisingly sophisticated as compared with other early instruments, and may represent early examples of Keplerian telescopes. About fifteen additional scientific and measuring devices can be noticed in the paintings. Of particular interest are an astrolabe, an armillary sphere, a nocturnal, a theodolite, a reduction compass, a Mordente's compass, and a gunner's rule. Most of these instruments can be associated with Michiel Coignet, instrument maker and cosmographer at the Court of Albert VII of Hapsburg.

## Acknowledgments

Thanks to Chiara Doz and Laura Abrami for valuable bibliographical help, and to Gabriella Schiulaz for language reading.